\documentclass[a4paper]{llncs}
\usepackage[]{sosy-paper}
\usepackage{graphicx}
\usepackage{subcaption}

\usepackage{tikz}
\tikzset{every picture/.style={/utils/exec={\sffamily}}}
\usepackage{amssymb}
\usetikzlibrary{arrows,positioning,shapes,arrows,decorations,backgrounds,shadows}
\usepackage{wrapfig}

\usepackage{tcolorbox}
\renewcommand{\fbox}[1]{\tcbox[on line,colback=black!15,colframe=black!15,left=1mm,right=1mm,top=0mm,bottom=0mm,boxsep=0mm,bottomrule=0mm]{#1}}

\tikzstyle{ca}=[draw,circle,fill=black!5,rounded corners]  % states
\tikzstyle{caf}=[draw,double, circle,fill=black!5,rounded corners] % accepting states
\tikzstyle{leer} = [rectangle,node distance=.6cm]              % can be used for placement
\tikzstyle{cloud} = [draw, ellipse, draw=blue!30!black, fill=blue!30, node distance=3cm, minimum height=2em]
\tikzstyle{circle-P} = [draw, circle, draw=blue!30!black, fill=red!30, node distance=3cm, minimum height=3.5em]
\tikzstyle{circle-phi} = [draw, circle, draw=blue!30!black, fill=blue!60, node distance=3cm, minimum height=3.5em]
\tikzstyle{circle-psi} = [draw, circle, draw=blue!30!black, fill=green!60, node distance=3cm, minimum height=3.5em]
\tikzstyle{circle-r} = [draw, circle, draw=blue!30!black, fill=red!70, node distance=3cm, minimum height=3.5em]
\tikzstyle{circle-w} = [draw, circle, draw=blue!30!black, fill=yellow!60, node distance=3cm, minimum height=3.5em]
\tikzstyle{circle-t} = [draw, circle, draw=blue!30!black, fill=gray!60, node distance=3cm, minimum height=3.5em]

\tikzset{>=latex}
\tikzstyle{ent}=[draw, fill=blue!20, text width=5em,
    text centered,minimum width=6em, minimum height=3em, rounded corners, drop shadow]

\usepackage{tikz-uml}

\usepackage{lmodern}

\usepackage{hyperref}
\hypersetup{
    colorlinks,
    linkcolor={red!50!black},
    citecolor={blue!50!black},
    urlcolor={blue!80!black}
}

\newcommand{\Inv}{\mathit{Inv}\xspace}

\newcommand{\pth}{\mathit{paths}\xspace}

\newcommand{\x}{\chi\xspace}
\newcommand{\prog}{\textit{p}\xspace}

\newcommand{\edges}{{\mathcal G}} 

\renewcommand{\true}{\textsc{true}\xspace} 
\renewcommand{\false}{\textsc{false}\xspace} 
\renewcommand{\unknown}{\textsc{unknown}\xspace}

\newcommand{\bspec}{\varphi_b\xspace}
\newcommand{\tspec}{\varphi_t\xspace}
\newcommand{\pred}{\gamma\xspace}

\newcommand{\Bspec}{\Phi_b\xspace}
\newcommand{\Tspec}{\Phi_t\xspace}
\newcommand{\Pred}{\Gamma\xspace}

\newcommand{\linefill}{\cleaders\hbox{$\smash{\mkern-2mu\mathord-\mkern-2mu}$}\hfill\vphantom{\lower1pt\hbox{$\rightarrow$}}}

\newcommand{\transi}[2]{\mathrel{\lower1pt\hbox{$\mathrel-_{\vphantom{#2}}\mkern-8mu\stackrel{#1}{\linefill_{\vphantom{#2}}}\mkern-11mu\rightarrow_{#2}$}}}
\newcommand{\trans}[1]{\transi{#1}{{}}}

\newif\iftechreport
\techreportfalse

\techreporttrue

\newcommand{\myvspace}[1]{\vspace{#1}}
\iftechreport
  \renewcommand{\myvspace}[1]{}
\fi

\newcommand{\mypaperkeywords}{
Cooperative Verification,
Software Verification,
Conditional Model Checking,
Verification Witness,
Exchange Format,
Partial Verification,
Reducer,
Execution Report,
Tool Combination
}

\title{Verification Artifacts in Cooperative Verification:\\
\large Survey and Unifying Component Framework}
\author{Dirk Beyer and Heike Wehrheim}

\begin{document}

% Redefine for title page
\author{Dirk Beyer$^1$%
~and Heike Wehrheim$^2$%
\vspace{1ex}}
\institute{%
$^1$\,%
LMU Munich, Germany \quad
$^2$\,%
Paderborn University, Germany}

\maketitle

\begin{abstract}
The goal of \emph{cooperative} verification is to combine verification approaches
in such a way that they work together to verify a system model.
In particular, cooperative verifiers
\emph{provide} exchangeable information (verification artifacts) \emph{to} other verifiers or
\emph{consume} such information \emph{from} other verifiers
with the goal of increasing the overall effectiveness and efficiency of the verification process.

This paper first gives an overview over approaches
for leveraging strengths of different techniques, algorithms, and tools
in order to increase the power and abilities of the state of the art in
software verification.

Second, we specifically outline cooperative verification approaches and discuss
their employed verification artifacts. We formalize all artifacts in a uniform way, 
thereby fixing their semantics and providing verifiers with a precise meaning of 
the exchanged information. 

\end{abstract}


%\newpage
\section{Introduction}

%oberbegriff: combination, darunter: 
%basic approaches, conceptual integrations, portfolio, cooperative verification

The area of software verification studies methods and 
constructs tools for automatically
proving program properties. 
The recent past has seen an enormous improvement in this area,
in particular with respect to scalability, precision,
and the handling of different programming-language features. 
Today's software-verification tools employ a variety of different techniques, 
ranging from data-flow analysis~\cite{Kildall}
over symbolic execution~\cite{King76}
to SAT-based approaches~\cite{BMC,AlgorithmComparison-JAR}. 
As all these techniques have their particular strengths and weaknesses, 
a number of tools \emph{tightly} integrate different ---usually two--- approaches
into one tool (see~\cite{HBMC-dataflow} for an overview). 
For instance, the integration of techniques that
under- and over-approximate the state space of the program
is a frequent combination. 
Such combinations typically improve over pure approaches.  
However, such combinations also require new tool implementations for every 
additional integration of techniques.
Portfolio combinations \emph{loosely} integrate different tools:
There is no communication between the approaches and the resulting combination
can be composed from off-the-shelf components.
Algorithm selection combines different approaches into one by
first analyzing the input problem and then choosing the
approach that will most likely (according to some heuristics) succeed.

In contrast to these extremely tight or extremely loose combinations, 
\emph{cooperative verification} is a combination of approaches 
that \emph{cooperate}, that is, work together to achieve the verification goal, 
but leave the existing tools (mostly) untouched. 
Cooperative verifiers \emph{communicate} with each other
in order to maximize the common strength, in particular,
by exchanging information about intermediate results. 
In a framework for cooperative verification, the integration of a new 
technique requires some implementation 
to make it understand the communication, viz.~be able to use intermediate results, 
but it can avoid a new re-implementation of the combination
--- from the conceptual as well as from the practical viewpoint.
If the intermediate results come in a format already accepted by the tool 
(e.g. as a program), the tool can even be employed as is. 

In this paper, we provide a classification of verification approaches 
according to the interface and type of combination employed;
\iftechreport
\else
in the extended version of this article~\cite{CooperativeVerification-TR}
\fi
we briefly survey combination approaches,  
for portfolio, selection, cooperative, and conceptual combination of verification approaches.  
We then discuss a number of aspects relevant to cooperative verification,  
in particular its objectives and prerequisites. 

\section{Classification of Verification Approaches}

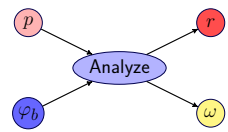
\begin{wrapfigure}{r}{38mm}
  \centering
  \vspace{-8mm}
  \scalebox{0.3}{
	%Verification
	%\begin{tikzpicture}[node distance = 10 mm and 10 mm,> = stealth']
	%\node (a) [cloud] at (-10,0) {\LARGE Verifier};
	%\node (p) [anchor=south east,inner sep=0] at (-14,0.5) {\includegraphics[width=.15\textwidth]{sheet}};
	%\node (o) [anchor=south east,inner sep=0] at (-14,-2.5) {\includegraphics[width=.15\textwidth]{sheet}};
	%\node (m) [text width=3cm,align=center] at (-14.8,.1) {\Large Program P};
	%\node (n) [text width=3.2cm,align=center] at (-15,-2.9) {\Large Specification $\varphi$};
	%\node (j) [anchor=south east, text width=5cm, align=center] at (-4,1) {\Large \textsc{True} ($P\vDash\varphi$)};
	%\node (k) [anchor=south east, text width=5cm, align=center] at (-4,-2) {\Large \textsc{False} ($P\nvDash\varphi$)};
	%\draw[->, line width=.5mm] (a) -- (j);
	%\draw[->, line width=.5mm] (a) -- (k);
	%\draw[->, line width=.5mm] (p) -- (a);
	%\draw[->, line width=.5mm] (o) -- (a);
	%\end{tikzpicture}
	%
    %Verifier
    \begin{tikzpicture}[node distance = 10 mm and 10 mm,> = stealth']
    \node (a) [cloud] at (-10,0) {\Huge Analyze};
    \node (p) [circle-P] at (-14,2) {\Huge $p$};
    \node (phi) [circle-phi] at (-14,-2) {\Huge $\varphi_b$};
    \node (r) [circle-r] at (-6,2) {\Huge $r$};
    \node (w) [circle-w] at (-6,-2) {\Huge $\omega$};
    \draw[->, line width=.5mm] (a) -- (r);
    \draw[->, line width=.5mm] (a) -- (w);
    \draw[->, line width=.5mm] (p) -- (a);
    \draw[->, line width=.5mm] (phi) -- (a);
    \end{tikzpicture}
  }
  \vspace{-1mm}
  \caption{Formal verification}
  \label{fig:verification-plain}
  \vspace{-8mm}
\end{wrapfigure}

In the following, we provide a classification of verification approaches 
according to their way of interfacing and combining verification components. 
By the term ``verification approach'' we understand 
an automatic or automatable formal method for solving {\em verification tasks}, 
i.e., for evaluating the proposition
``Program~$p$ satisfies behavioral specification~$\varphi_b$'' and returning
a result~$r$, which can be $\true~(p \models \varphi_b)$, $\false~(p \not\models \varphi_b)$, or $\unknown$,
and an (optional) witness~$\omega$, which contains proof hints,
as depicted in \cref{fig:verification-plain}.

\subsection{Overview over Interfaces}

\begin{wrapfigure}{r}{50mm}
  \vspace{-10mm}
  \scalebox{0.4}{
	%Output Interfacing
	\begin{tikzpicture}[node distance = 10 mm and 10 mm,> = stealth']
	\node (a) [cloud] at (0,0) {\LARGE Verifier};
	\node (p) [anchor=south west,inner sep=0] at (4,0) {\includegraphics[width=.2\textwidth]{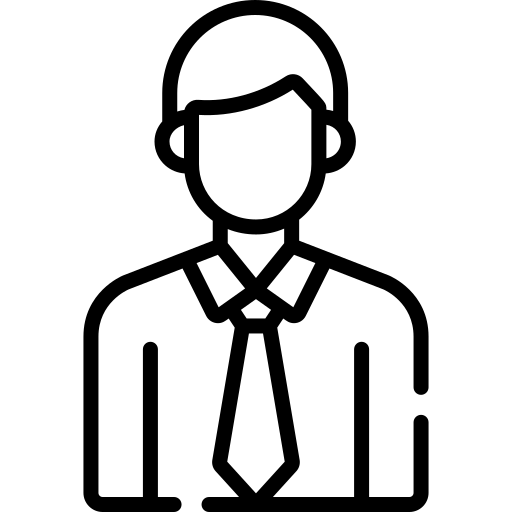}};
	\node (o) [anchor=south west,inner sep=0] at (4.1,-2) {\includegraphics[width=.2\textwidth]{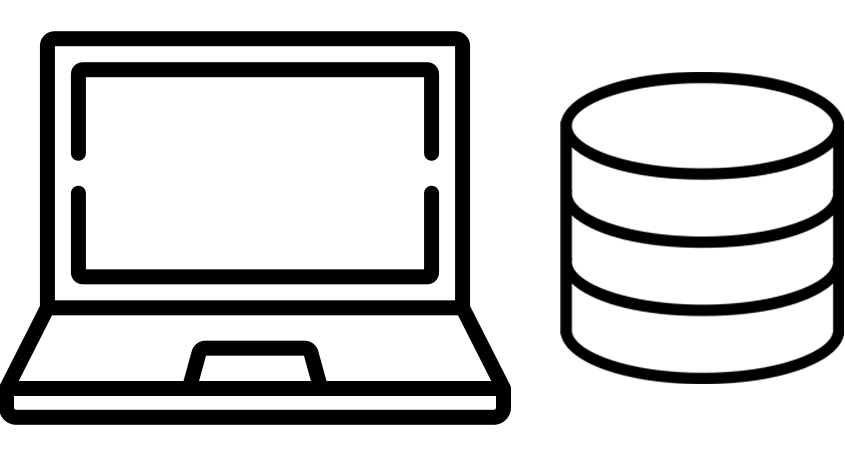}};
	\node (m) [text width=5cm, right = of p, align=center] at (5,1.0) {\Large Consumption\\by humans};
	\node (n) [text width=5cm, right = of o, align=center] at (5,-1.2) {\Large Consumption\\by machines};
	\draw[->, line width=.5mm] (a) -- (p);
	\draw[->, line width=.5mm] (a) -- (o);
	\end{tikzpicture}
  }
  \caption{Output Interfaces}
  \label{fig:interfacing-output}
  \vspace{-8mm}
\end{wrapfigure}
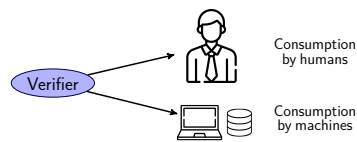
\subsubsection{Output}
The goal of a verification tool is to solve a verification task and
to deliver the computed results to either a verification engineer for manual inspection
or to a machine for further automated processing (\cref{fig:interfacing-output}).
Depending on how the results are consumed (by human or by machine),
the tool needs to use different formats. 

While researchers mainly concentrated on improving the (internal) verification algorithms
during the past two decades, it is understood since recently that it is (at least) equally
important to provide not only \true/\false answers,
but more details about the reasoning and the process of the verification. 

\inlineheadingit{Human}
Almost all verification tools provide some kind of statistics to the user,
for example, about the number of iterations, of number of proof facts, or consumed resources.
\emph{Execution reports}~\cite{ExecutionReports}
present an underapproximation of the successfully verified state space to the user. 
There are also approaches to support interactive inspection of verification results,
e.g., by visualization of error paths~\cite{KLEVER} and
verification-aided debugging~\cite{VerificationAidedDebugging}.

\inlineheadingit{Machine}
In order to make it possible to validate verification results in an automated way,
\emph{verification witnesses} were introduced~\cite{ReuseVerificationResultsSPIN,SVCOMP15},
a machine-readable exchange format (XML based).
Verification witnesses make it possible to independently re-verify the program
based on knowledge that another verifier has produced.
This can increase trust in the results, can spread the risk of verification errors,
and can help making internal knowledge from the verification engine accessible for the user
(error paths, program invariants).
\emph{Violation witnesses}~\cite{Witnesses} enhance the answer \false by a description of
the state space that contains an error path (a program path that violates the specification),
while \emph{correctness witnesses}~\cite{CorrectnessWitnesses}
enhance the answer \true by a description of program invariants
that are helpful to prove that the program satisfies the specification.
It is known since 15 years that test cases can be derived from error paths~\cite{BLAST-test,PathFinder-test},
but this approach was rarely used in practice and only since recently it is possible to output
and exchange this kind of information via a standard format.

While the previous approaches, as the name indicates, witness the verification result,
it is also important to make intermediate results and partial results accessible
to further processing.
\emph{Conditional model checking}~\cite{ConditionalModelChecking} reads as input and writes as output
a description of the already verified state space.
That is, a conditional verifier outputs a condition that describes the work already done,
i.e, the parts of the state space that are already verified.
Another kind of intermediate output for machines to later reuse it is
the \emph{abstraction precision}~\cite{PrecisionReuse,CPAplus}.
In CEGAR-based approaches~\cite{ClarkeCEGAR} an abstract model is automatically constructed
by finding abstraction facts in refinement steps which are added to the precision of the analysis
(the more abstraction facts are added to the precision, the finer the abstract model).
Full abstract models can be used as certificate of correctness~\cite{jakobs2014certification}
or in order to speed up later verification runs for different versions of the same program
during regression verification~\cite{ExtremeModelChecking}.

\begin{wrapfigure}{r}{50mm}
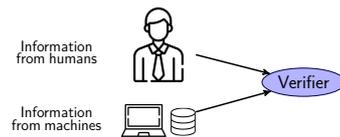

  \vspace{-8mm}
  \scalebox{0.4}{
	%Input Interfacing
	\begin{tikzpicture}[node distance = 10 mm and 10 mm,> = stealth']
	\node (a) [cloud] at (0,0) {\LARGE Verifier};
	\node (p) [anchor=south west,inner sep=0] at (-6,0) {\includegraphics[width=.2\textwidth]{human}};
	\node (o) [anchor=south west,inner sep=0] at (-6,-2) {\includegraphics[width=.2\textwidth]{computer}};
	\node (m) [text width=5cm, left = of p, align=center] at (-4.5,1.0) {\Large Information\\from humans};
	\node (n) [text width=5cm, left = of o, align=center] at (-4.5,-1.2) {\Large Information\\from machines};
	\draw[->, line width=.5mm] (p) -- (a);
	\draw[->, line width=.5mm] (o) -- (a);
	\end{tikzpicture}
  }
  \vspace{-6mm}
  \caption{Input Interfaces}
  \label{fig:interfacing-input}
  \vspace{-10mm}
\end{wrapfigure}
\subsubsection{Input}
Similar to the output, there are different interfaces for the kind of input that is given to the
verification tools, some from users, some from machines, see \cref{fig:interfacing-input}.

\inlineheadingit{Human}
From the very beginning of programming, assertions were added to programs~\cite{Turing49}
in order to make it easier to prove correctness.
Nowadays, assertions, invariants, pre- and post-conditions, are annotated in programs
in a way that machines (interactive verifiers) can read.
There are several languages and tools that support this,
and a nice overview over such tools and their application opportunities
are given in the annual competition on interactive software verification VerifyThis~\cite{VerifyThis19}.

There were also attempts to support the splitting of specifications and programs
into modular parts, in order to make the verification task for the model checkers easier,
such as in the \blast query language~\cite{BLAST-query,SeryBLAST}.
Last not least, and this is one of the most difficult parts, each verifier
expects a set of parameters that the user has to set, in order to choose how the verifier
should solve its task.
However, finding the right parameters is a challenging task, which could use tool support itself
(such as SMAC~\cite{SMAC} or Tuner~\cite{Tuner}).

\inlineheadingit{Machine}
A classic approach to make additional information available to a tool is by transforming
the original input, e.g., by simplification or enhancement.
The advantage is that there is no additional input
(no extra parser, no need to implement additional features).
For example, the first software model checkers did not have a
specification language, but the specification was weaved into the program
in a preprocessing step
(as was done for the \slam~\cite{SLAM} specification language \tool{Slic}~\cite{SLIC}
and the \blast~\cite{BLAST} query language~\cite{BLAST-query}).
Even programs were made simpler~\cite{CIL}.

Verification witnesses and conditions were discussed already above as example
implementations for output interfaces.
Verification witnesses can be taken as input by validation tools that re-establish the
verification result using independent technology.
Also, the error path described by the violation witness can be replayed and a test case
can be derived from the path constraints along the found error path\cite{ExecutionBasedValidation}.

Conditional model checking is not widespread yet because it was considered
difficult to extend a verifier such that it understands conditions as input
and reduces the state space accordingly before running the verification engine.
This problem was solved by the reducer-based construction of conditional verifiers:
\emph{Reducers}~\cite{Reducer,CzechConditionalModelChecking}
can be used to construct (without implementation effort)
conditional model checkers from off-the-shelf verifiers that do not understand conditions themselves,
by reducing the original input program to a residual program that contains
all the behavior that is not yet covered by the condition and removes as much as possible
from the already-verified state space.

\subsection{Overview over Combinations} 
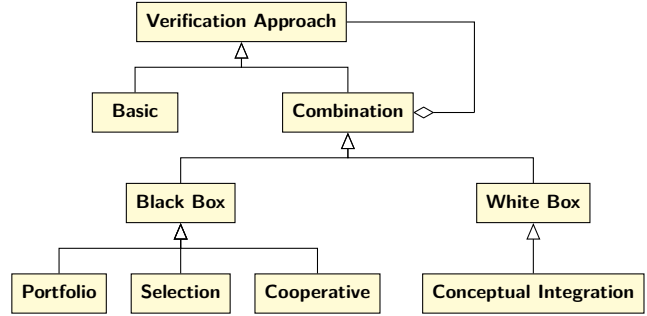
\begin{figure}[t] 
  \begin{center} 
    \sffamily
      \begin{tikzpicture} [
        scale = 0.8,
        transform shape,
        every node/.style = {
          align = center,
          font = \footnotesize,
        }
      ]
    %\begin{umlpackage}[x=0, y=0, transform shape = false]{package-name}

    \umlsimpleclass[x=0,y=0]{Verification Approach}
    \umlsimpleclass[x=-1.75,y=-1.5]{Basic}
    \umlsimpleclass[x=1.75,y=-1.5]{Combination}
    \umlsimpleclass[x=-1,y=-3]{Black Box}
    \umlsimpleclass[x=4.8,y=-3]{White Box}
    \umlsimpleclass[x=-3,y=-4.5]{Portfolio}
    \umlsimpleclass[x=-1,y=-4.5]{Selection}
    \umlsimpleclass[x=1.25,y=-4.5]{Cooperative}
    \umlsimpleclass[x=4.8,y=-4.5]{Conceptual Integration}

    \umlinherit[geometry=|-|]{Basic}{Verification Approach}
    \umlinherit[geometry=|-|]{Combination}{Verification Approach}
    \umlinherit[geometry=|-|]{White Box}{Combination}
    \umlinherit[geometry=|-|]{Black Box}{Combination}
    \umlinherit{Conceptual Integration}{White Box}
    \umlinherit[geometry=|-|]{Portfolio}{Black Box}
    \umlinherit[geometry=|-|]{Selection}{Black Box}
    \umlinherit[geometry=|-|]{Cooperative}{Black Box}

    \umlemptyclass[right=1cm of Combination,coordinate,opacity=0]{M}
    \umlaggreg[geometry=--]{Combination}{M};
    \umlassoc[geometry=|-]{M}{Verification Approach};

    %\end{umlpackage} 
  \end{tikzpicture}
  \end{center}
  \myvspace{-6mm}
  \caption{Hierarchy of verification approaches (using UML notation)}
  \label{fig:combinations}
  \myvspace{-8mm}
\end{figure}

In the early days of automatic program verification, 
tools implemented a single technique for verification 
(e.g., explicit enumeration of state space 
or data-flow analysis using a fixed abstract domain). 
In our classification (see Fig.~\ref{fig:combinations}) these are represented as \textbf{Basic}. 
Later, the tools implementing these techniques 
were considerably generalized, 
for instance by allowing abstract domains to be 
flexibly set via tool parameters. 
Still, during one verification run a single basic technique 
was employed. 

It soon turned out  
that a single verification technique may work well 
for some verification tasks, but fail for others. 
This immediately triggered the application of \textbf{Combination} techniques,
in order to benefit from the different strengths.
Combinations can come in two sorts:  
A combination either treats techniques or tools 
as \textbf{Black Box} objects
and runs them (mainly) as they are without implementation-specific integrations
for which it matters what's inside the box, 
or a combination views a component as \textbf{White Box}, 
conceptually integrating two or more techniques within a new tool.
We distinguish three forms of black-box combinations, without and with communication,
and classify all white-box approaches into one category.

\begin{wrapfigure}{r}{50mm}
  \centering
  \myvspace{-1mm}
  \scalebox{0.4}{
	%Portfolio-sequential
	\begin{tikzpicture}[node distance = 10 mm and 10 mm,> = stealth']
	\coordinate[] (z) at (0,0);
	\coordinate[] (y) at (6,0);
	\coordinate[] (x) at (4.5,0);
	\node (a) [cloud] at (2.5,0)   {\Large Verifier 1};
	\node (b) [cloud] at (8,0)    {\Large Verifier n};
	\draw[<-, line width=.5mm] (z) -- +(180:2cm);
	\draw[->, line width=.5mm] (z) -- (a);
	\draw[->, line width=.5mm] (a) -- (x);
	\draw[->, line width=.5mm] (y) -- (b);
	\path (x) -- node[auto=false]{\LARGE \ldots} (y);
	\draw[line width=.2mm] (5,0) ellipse (5cm and 1.5cm);
	\end{tikzpicture}
  }
  \vspace{0mm}\\
  \scalebox{0.4}{
	%Portfolio-parallel
	\begin{tikzpicture}[node distance = 10 mm and 10 mm,> = stealth']
	\coordinate[] (z) at (0,0);
	\node (a) [cloud, below right=of z,yshift=.7cm]    {\Large Verifier n};
	\node (b) [cloud, above right=of z,yshift=-.7cm]    {\Large Verifier 1};
	\draw[<-, line width=.5mm] (z) -- +(180:2cm);
	\draw[->, line width=.5mm] (z) -- (a);
	\draw[->, line width=.5mm] (z) -- (b);
	\path (b) -- node[auto=false]{\LARGE \ldots} (a);
	\draw[line width=.2mm] (3.2,0) ellipse (4cm and 2.5cm);
	\end{tikzpicture}
  }
  \caption{Portfolio approaches\\ (top: sequential, bottom: parallel)}
  \label{fig:portfolio}
  \myvspace{-8mm}
\end{wrapfigure}
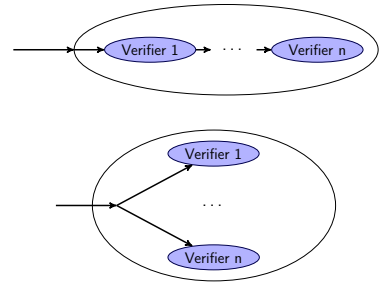
\textbf{Portfolio} combinations are motivated by the
portfolio idea from economics~\cite{Portfolio},
which is a means of distribution of risk:
if one investment (here: of computational resources in a certain technique) fails,
there are other investments (techniques) that will be successful.
A portfolio combination has a number of approaches available,  
and on a given verification task 
executes the approaches in a fixed order sequentially (\cref{fig:portfolio}, top),
or runs all approaches in parallel (\cref{fig:portfolio}, bottom).
The overall approach terminates if one component analysis was successful in
obtaining the result. 
The big advantage of this approach is that it requires no knowledge about the
components and there is almost no effort for implementing the combination.
Therefore, we placed this most loosely coupled approach
on the very left in the bottom row of our Fig.~\ref{fig:combinations}.
The big disadvantage of portfolio approaches
is that the resources invested on unsuccessful tools or approaches are lost.

\begin{figure}[t]
  \centering
  \scalebox{0.7}{
	%Selection
	\begin{tikzpicture}[node distance = 10 mm and 10 mm,> = stealth']
	\coordinate[] (z) at (0,0);
	\node (a) [cloud, align=center] at (2,0) {Selection-\\ Model\\Extraction};
	\node (b) [cloud] at (7.5,0)    {Selector};
	\node (c) [cloud] at (7,-2)    {Verifier selected};
	\node (d) [cloud] at (5,1.5)    {Verifier 1};
	\node (e) [cloud] at (8.5,1.5)    {Verifier n};
	\node (f) [draw,align=center] at (5,0)    {Selection\\ Model};
	\draw[<-, line width=.5mm] (z) -- +(180:2cm);
	\draw[->, line width=.5mm] (z) -- (a);
	\draw[->, line width=.5mm] (a) -- (f);
	\draw[->, line width=.5mm] (f) -- (b);
	\draw[->, line width=.5mm] (d) -- (b);
	\draw[->, line width=.5mm] (e) -- (b);
	\draw[->, line width=.5mm] (b) -- (c);
	\draw[->, line width=.5mm] (z) to [out=-50,in=-180] (c); %[out=-50,in=-150]
	\path (d) -- node[auto=false]{\LARGE \ldots} (e);
	\draw[line width=.2mm] (5.5,0) ellipse (5.5cm and 2.7cm);
	\end{tikzpicture}
  }
  \myvspace{-3mm}
  \caption{Algorithm selection}
  \label{fig:alg-selection}
  \myvspace{-8mm}
\end{figure}
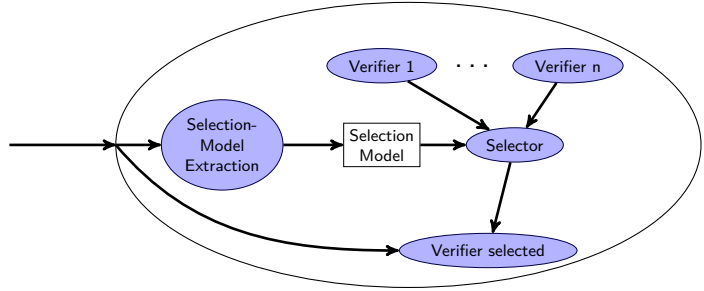

Algorithm \textbf{Selection}~\cite{AlgorithmSelection} is a solution to the problem of wasted resources
of portfolio approaches:
Algorithm-selection approaches have a number of approaches available,  
and on a given verification task choose one and execute it
(\cref{fig:alg-selection}).
That is, before starting an approach, a selection model is extracted from the input,
based on which a selector function predicts which approach would be best,
and only the potentially best approach is selected and executed.
This requires some knowledge about the (black box) characterization of the components,
but does not require any change of the implementation of the components.

Portfolio and selection approaches run the component tools independently from each other, 
without any form of information exchange between the approaches.
The goal of combining strengths of different approaches
and at the same time avoiding to waste resources
inspired the development of {\em cooperative} combinations 
of verification approaches.

\textbf{Cooperation} approaches % overcome this limitation and
enable the possibility of solving the problem \emph{together}.
Typically, tools exchange intermediate results
(e.g., the state space which has already been searched)
in order to achieve a division of labor. 
Such cooperative combinations range from two or more 
basic techniques running in parallel and combining 
the information obtained for certain program locations 
(e.g., combining partial verification results to proof witnesses~\cite{CombiningPartialResults})
to approaches executing different tools in turns
with each specializing to specific tasks
(e.g.,~a testing tool trying to achieve coverage 
together with a model checker constructing 
counter examples to non-reachability~\cite{AbstractionDrivenConcolicTesting}). 

\textbf{Conceptual Integration} is most intensively coupled approach and therefore
put on the very right end of the bottom row in our Fig.~\ref{fig:combinations}.
The components are not communicating via clear interfaces,
but are tightly integrated and exchange data structures via procedure calls
and not via interfaces that could be externalized~\cite{HBMC-dataflow}.

\iftechreport
  In the following subsections, we describe
\else
  The extended version of this article~\cite{CooperativeVerification-TR} describes
\fi
some forms of non-cooperative verification approaches in more detail.
In the next section we explain some examples for cooperative verification approaches.

\nocite{Portfolio}
\nocite{SLAM2}
\nocite{MUX}
\nocite{YOGI}
\nocite{SVCOMP13}
\nocite{CPACHECKERSEQCOM-COMP13}
\nocite{ABE}
\nocite{PREDATOR-COMP15}
\nocite{AlgorithmSelection}
\nocite{StrategySelection}
\nocite{MUX}
\nocite{PortfolioVerification,PortfolioVerificationCAV15}
\nocite{PredictingRankings}
\nocite{DBLP:conf/popl/GodefroidNRT10}
\nocite{DBLP:journals/tse/BeckmanNRSTT10}
\nocite{Synergy}
\nocite{CPACHECKER}
\nocite{CousotPOPL79}
\nocite{GulwaniTiwariPLDI06,CodishPEPM93,LernerPOPL02,PredicatedLattice,LazyShapeAnalysis}
\nocite{HBMC-dataflow}
\nocite{HBMC-testing}

\iftechreport
\subsection{Examples for Portfolio Combinations}

While it seems obvious that combinations of verification techniques 
have a large potential, 
for software verification the topic is not yet systematically investigated,
while it is used in other areas since many years~\cite{Portfolio}.

\inlineheadingit{Sequential Combinations}
Examples of sequential combinations are \sdv and \cpachecker.
The static driver verification (SDV)~\cite{SLAM2} tool chain at Microsoft
used a sequential combination (described in~\cite{MUX}) which first
runs Corral~\cite{CORRAL} for up to \SI{1400}{\second} and then Yogi~\cite{YOGI}.
\cpachecker~\cite{CPACHECKER} won the competition on software verification 2013 (SV-COMP'13, \cite{SVCOMP13})
using a sequential combination~\cite{CPACHECKERSEQCOM-COMP13}
that started with explicit-state model checking for up to~\SI{100}{\second}
and then switched to a predicate analysis~\cite{ABE}.

\inlineheadingit{Parallel Combinations}
Examples of parallel combinations are the
verifiers \ufo~\cite{UFO-COMP13} and \predatorhp~\cite{PREDATOR-COMP15},
which start several different strategies simultaneously and take the result
from the approach that terminates first.

\subsection{Examples for Algorithm Selection}
Algorithm selection~\cite{AlgorithmSelection}
first extracts a \emph{selection model} from the input.
In the case of software verification, the input is the verification task (program and its specification).
The selection model describes some characteristics of the verification task,
for example, feature vectors (measurement values for certain measures that map verification tasks to values).
Based on the selection model, the \emph{strategy selector} chooses one strategy from a
set of given verification strategies.

\inlineheadingit{Approaches without Machine Learning}
Strategy selection can be very simple and yet effective.
For example, a recent work has shown that it is possible with a few boolean features
to effectively improve the overall verification progress~\cite{StrategySelection}.
The disadvantage is that the strategy selector needs to be explicitly defined by the developer or user.
This leads to approaches that use machine learning, in order to automatically learn the
strategy selector from training instances.

\inlineheadingit{Machine-Learning-Based Approaches}
The technique~MUX~\cite{MUX} can be used to synthesize a strategy selector
for a set of features of the input program and a given number of strategies.
The strategies are verification tools in this case, and the feature values
are statically extracted from the source code of the input program.
Later, a technique that uses more sophisticated features
was proposed~\cite{PortfolioVerification,PortfolioVerificationCAV15}.
While the above techniques use explicit features (defined by measures on the source code),
a more recently developed technique~\cite{PredictingRankings}
leaves it up to the machine learning to obtain insights from the input program.
The advantage is that there is no need to define the features:
the learner is given the control-flow graph, the data-dependency graph, and the abstract syntax tree,
and automatically derives internally the characteristics that it needs.
Also, the technique predicts a ranking, that is, the strategy selector is not a
function that maps verification tasks to a strategy, but to a sequence of strategies.

\subsection{Examples for Conceptual Integrations} 

Conceptual integrations tightly combine two or more approaches 
into a new tool, typically re-implementing the basic techniques.
A frequent combination of this type is integrating 
an overapproximating (static) may-analysis with 
an underapproximating (dynamic) must-analysis. 
The tool SMASH~\cite{DBLP:conf/popl/GodefroidNRT10} 
at the same time maintains an
over and an under approximation 
of the state space of programs. 
Building on the same idea, Beckman et al.~\cite{DBLP:journals/tse/BeckmanNRSTT10} 
(tool Yogi, first proposal of the algorithm under name \textsc{Synergy} in~\cite{Synergy}) 
in addition specifically employs testing to 
derive alias information which is costly to precisely compute by a static analysis. 

A second form of conceptual integration is offered 
by tools running different analysis in parallel in a form of ``product'' construction. 
A prototypical example is the tool \cpachecker~\cite{CPACHECKER} with the 
possibility of specifying and running {\em composite} analyses. 
A composite analysis could for instance combine two sorts of data-flow analyses 
(e.g., an interval analysis and an available-expression analysis). 
The analyses are then jointly run and jointly derive analysis information 
for program locations.
The same idea was classically hard-coded as
reduced product~\cite{CousotPOPL79}
and further improved~\cite{GulwaniTiwariPLDI06,CodishPEPM93,LernerPOPL02,PredicatedLattice,LazyShapeAnalysis}.

All those combinations have in common that they exchange information,
but they are more intertwinned, hardcoded combinations,
rather than interface-based combinations.
More approaches are described in the Handbook on Model Checking,
in the chapters on combining model checking with data-flow analysis~\cite{HBMC-dataflow},
with deductive verification~\cite{HBMC-deduction}, and with testing~\cite{HBMC-testing}.

\fi

\section{Cooperative Verification Approaches}

In the following, we discuss approaches for cooperative verification,
structured according to the kind of information objects which are exchanged,
and then explain a few applications and their effects.

\subsection{Exchangeable Objects for Communication and Information Transfer}

We now classify the approaches for cooperative verification according to the kinds of communication interfaces that they use.
While our text always refers to software verification for concrete examples,
cooperative verification is in no way limited to software.

\inlineheadingbf{Conditions and Residual Programs}
Conditional model checking~(CMC)~\cite{ConditionalModelChecking}
means to produce a condition as output that describes the state-space that was successfully verified.
The (temporal) condition can be represented as an automaton.
This information can be passed on to another verifier as input,
instructing this verifier to not verify again those parts of the state space that are covered by the condition.
Using a \emph{reducer}~\cite{Reducer}, a program can be reduced to those parts of its state space
that still has to be verified; the result is called \emph{residual program}.
Symbiotic~\cite{SYMBIOTIC5-SVCOMP18} can be seen as reducer-based cooperation
(slicer + \tool{Klee}).

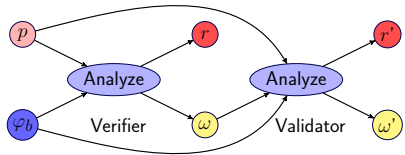
\begin{figure}[t]
  \myvspace{-4mm}
  \centering
  \scalebox{0.3}{
	%Witness-Based Results Validation
    %Verifier
    \begin{tikzpicture}[node distance = 10 mm and 10 mm,> = stealth']
    \node (a) [cloud] at (-10,0) {\Huge Analyze};
    \node (l) [text width=5cm] at (-8.5,-2) {\Huge Verifier};
    \node (p) [circle-P] at (-14,2) {\Huge $p$};
    \node (phi) [circle-phi] at (-14,-2) {\Huge $\bspec$};
    \node (r) [circle-r] at (-6,2) {\Huge $r$};
    \node (w) [circle-w] at (-6,-2) {\Huge $\omega$};
    \draw[->, line width=.5mm] (a) -- (r);
    \draw[->, line width=.5mm] (a) -- (w);
    \draw[->, line width=.5mm] (p) -- (a);
    \draw[->, line width=.5mm] (phi) -- (a);
    \node (za) [cloud] at (-2,0) {\Huge Analyze};
    \node (zl) [text width=5cm] at (-0.4,-2) {\Huge Validator};
    \node (zr) [circle-r] at (2,2) {\Huge $r$'};
    \node (zw2) [circle-w] at (2,-2) {\Huge $\omega$'};
    \draw[->, line width=.5mm] (za) -- (zr);
    \draw[->, line width=.5mm] (za) -- (zw2);
    \draw[->, line width=.5mm] (p) to [out=15,in=120] (za);
    \draw[->, line width=.5mm] (phi) to [out=-15,in=-120] (za);
    \draw[->, line width=.5mm] (w) -- (za);
    \end{tikzpicture}
  }
  \vspace{-4mm}
  \caption{Witness-based results validation}
  \label{fig:verification-witness-based}
  \vspace{-6mm}
\end{figure}

\inlineheadingbf{Witnesses}
Exchangeable witnesses serve as envelopes for error paths and invariants in a way that makes it possible to
exchange the information between different tools.
A~\emph{violation witness}~\cite{Witnesses,ReuseVerificationResultsSPIN,SVCOMP15} explains the specification violation,
by describing a path through the program that violates the specification.
A~\emph{correctness witness}~\cite{CorrectnessWitnesses} explains why the program satisfied the specification,
by describing invariants that are useful to have in a correctness proof.
\Cref{fig:verification-witness-based} illustrates the process:
The first analyzer verifies the program~$p$ according to specification~$\bspec$,
and produces a result~$r$ and a witness~$\omega$.
The second analyzer \mbox{(re-)verifies} the same program and specification
using information from the witness.
If the result~$r$ matches the result~$r$', then the result is confirmed.

\inlineheadingbf{Precisions}
Verification approaches that are based on counterexample-guided abstraction refinement~(CEGAR)~\cite{ClarkeCEGAR}
iteratively construct an abstract model of the system.
The ``abstraction facts'' that define the level of abstraction are often
formalized and expressed as \emph{precision}~\cite{PrecisionReuse,ReuseVerificationResultsSPIN,CPAplus}.
The precision can be exported as output, such that later verification runs can start from such a given
definition of the abstraction level.

\inlineheadingbf{Abstract States / Certificates}
Extreme model checking~\cite{ExtremeModelChecking} dumps the abstract reachability graph (ARG) to a file when the verification process terminates. 
Configurable certificates~\cite{jakobs2014certification} are sets of abstract states that cover all reachable
states of the system.
ARGs and configurable certificates can be used by a different verifier to check its validity (completeness and soundness).

\inlineheadingbf{Path Programs and Path Invariants}
Path programs~\cite{PathInvariants} are programs (for example, written in the same programming language as the input program)
that were invented to incorporate external invariant generators into CEGAR-based approaches
and are produced after a verifier has found an infeasible error path (often called infeasible counterexample).
The path program contains that path in question, but usually also longer paths that use the same program operations,
that is, unrollings of a certain loop.
The path program can now be given to a tool for invariant synthesis (e.g.,~\cite{InvariantSynthesisCombinedTheories})
in order to obtain \emph{path invariants}~\cite{PathInvariants},
which are invariants for the whole path program, but in particular also for the original path.
The path invariants can then be fed back into the CEGAR-based approach that was encountering the original path.

\subsection{Objectives and Applications}
Having exchangeable objects about (partial) verification results~\cite{PartialVerification} available is important to overcome
a variety of practical problems.
In the following, we highlight a few of the objectives and applications that we can aim for.

\inlineheadingbf{Improvement of Effectiveness and Efficiency}
Storing intermediate results can be used to improve the effectiveness and efficiency of the verification process.

\inlineheadingit{Stateful Verification and Reuse}
Storing (exchangeable) objects that contain information about intermediate verification results
can be considered as a \emph{state} of the verification process,
i.e., making the verification process \emph{stateful}.

\emph{Precisions} that are stored and in later verification runs read and reused
can significantly improve the performance of regression verification~\cite{PrecisionReuse}.
The setup of this strategy is the following:
the first version of a module is verified and at the end, the precision is written to a file.
When the $i$-th version is verified, then the verifier reads the precision that the
verification run for version~$i-1$ has written, in order to save time discovering the
right abstraction level for the abstract model.

\emph{Configurable certificates}~\cite{jakobs2014certification} can reduce the validation time,
because the verifier that performs the validation of the certificate does ``only'' need to
check for the set of abstract states that all initial states are contained and
that the set is closed under successor transitions.

Also \emph{caching-based approaches} to improve the efficiency can be seen as a stateful
way of performing computation.
For example, Green~\cite{greenModelChecking} makes symbolic execution more efficient
by caching previous intermediate results.

\inlineheadingit{Stateless Verification and Parallelization}
The previous argument was based on having a state that contains the intermediate results.
It is also possible to speed up verification processes in a \emph{stateless} way.
The technique of conditional model checking is used to split programs into parts
that can be independently verified~\cite{StructurallyDefinedCMC}.

\inlineheadingbf{Improvement of Precision and Confidence}
\emph{Witness-based results validation}~\cite{Witnesses,CorrectnessWitnesses} can be used to increase the confidence in the
results of verification tools, because it is possible to 
take a witness-based results validator to ``replay'' the verification.
That is, for a violation witness, the validator tries to find and confirm the error path
that the witness describes,
and for a correctness witness, the validator tries to use the invariants in the witness to re-establish
the proof of correctness.

\emph{Execution-based results validation}~\cite{ExecutionBasedValidation}
extracts a test case from a violation witness and executes it,
in order to confirm that the specification violation is observable in the executed program as well.

%One of the first ideas to combine different tools was for eliminating
%false alarms: after the core verifier has found an error path,
%this error path is not immediately reported to the user, but first
%converted into a program again which is then verified by an external verifier,
%and only if that external tool reports an error path as well, then the
%alarm is shown as a result to the user.
%\footnote{An early version of \cpachecker~\cite{CPACHECKER} had constructed a path program~\cite{PathInvariants},
%dumped it to a file in C syntax, and then called \cbmc~\cite{CBMC} as external verifier for validation.
%Meanwhile, such an error-path check is a standard component in many verifiers.}

\inlineheadingbf{Explainability}
The existence and availability of exchangeable objects with information about the verification process
makes it possible to develop approaches for \emph{explaining} what the verification process did
and why the user should be more confident about the verification result.
There are preliminary results on explaining and visualizing counterexamples,
e.g., for SPIN models~\cite{CounterexampleExplanation} and for C programs~\cite{VerificationAidedDebugging},
but due to the exchangeable witness format, many more approaches are possible.

%\subsubsection{Execution Reports}
%

%\subsubsection{Coverage Information}
%

%\subsubsection{Test Cases}
%It is realatively easy to derive test values from an error path~\cite{BLAST-test,PathFinder-test}.

%\subsubsection{Cross-Technology Interaction}
%
%Construct an interactive theorem prover from a witness validator using invariants in correctness witnesses
%
%Test generators made from model checking technology

% Grant Proposal Dirk durchschauen nach Ideen

%%%dbeyer: We did not want to use this classification criteria anymore
%
%\subsection{Online vs. Offline}
%
%blackbox, communication over interface without knowing internals
%
%existing separate components
%
%Online: run simultaneously or directly in sequence, and exchange information 
%
%Offline: Saving intermediate results to a file and start another process,
%not time dependent, can be the other day

% !TeX root = main.tex
\section{Verification Artifacts} 

This section outlines a construction framework for cooperation. 
We study verification \emph{artifacts}, 
classify several verification tools as verification {\em actors}
according to their usage of artifacts, and
define the \emph{semantics} of some important artifacts.

\subsection{Artifacts of Verification Tools}
Verification artifacts are central to cooperation as they provide the means of information exchange.
A number of artifacts exist already, most notably of course the programs themselves.
We identified the following artifacts:
\begin{description}
\item[Program~$p$.]
   Defines the implemented behavior of the system.
   \textbf{Syntax}: C programming language (for example).
   We represent programs as control-flow automata in \cref{sec:semantics}.
\item[Behavior Specification $\bspec$.]
   Defines requirements that all executions of a given program have to satisfy,
   often as conjunction of several properties.
   \textbf{Syntax}: The competition SV-COMP established a minimal set of properties that
   participants of the competition have to support\,%
   \footnote{\url{https://sv-comp.sosy-lab.org/2019/rules.php}},
   which is based on LTL~\cite{HBMC-TemporalLogic}, but some tools also support monitor automata as specification.
   We represent properties by property automata in \cref{sec:semantics}.
\item[Test Specification $\tspec$.]
   Defines requirements that a given test suite has to sa\-tisfy.
   \textbf{Syntax}: The competition Test-Comp established a minimal set of coverage criteria that
   participants of the competition have to support\,%
   \footnote{\url{https://test-comp.sosy-lab.org/2019/rules.php}},
   which is based on FQL~\cite{FShell,FQL}, but some tools also offer parameters for hard-coded coverage criteria.
   We represent coverage criteria via  test-goal automata in \cref{sec:semantics}. 
\item[Verification Result $r$.]
   Verification tools return an evaluation of the statement
   ``Program~$p$ satisfies specification~$\bspec$.'' as answer,
   which is from the set~$\{\true, \false, \unknown\}$. 
\item[Witness $\omega$.]
  Verification witnesses are used to witness an outcome of a verification run, and 
  thus can come in the form of violation and correctness witnesses.
  \textbf{Syntax}: XML-based witness format\,%
  \footnote{\url{https://github.com/sosy-lab/sv-witnesses}}
  that is supported by all available validators of verification results.
\item[Test case $t$.]
  Defines a sequence of values for all calls of external functions, i.e., inputs for the program.
  \textbf{Syntax}: XML-based test-case format\,%
  \footnote{\url{https://gitlab.com/sosy-lab/software/test-format}}
  that is supported by all test-case generators that participate in Test-Comp.
\item[Condition $\psi$.]
  Defines the part of the program behavior that does not need to be further explored.
  For verification, $\psi$ describes the already verified parts.
  For testing, $\psi$ describes the parts of the program that are already covered by an existing test suite.
  \textbf{Syntax}: Condition automata using a notation similar to the \blast query language~\cite{BLAST-query}.
\end{description}
We use the corresponding capital letters to denote the types (i.e., sets of artifacts of a kind),
for example, the type~$P$ is the set of all C programs.
Many tools generate different forms of verification artifacts, 
but currently only very few understand more than the artifact ``program'' as input. 

\subsection{Classification of Verification Tools as Actors} 
Based on the identified artifacts, we classify existing tools according to their usage of artifacts
into three sorts of verification actors: 
\begin{description}
\setlength\itemsep{0em}
	\item[Analyzers.]    Produce new knowledge about programs, for example verification results or test suites.
	\item[Transformers.] Translate one artifact into another, in order to implement a certain feature or support cooperation.
	\item[Presenters.]   Prepare information from artifacts such that it can be presented in a human-readable form.
\end{description} 

To convey a better understanding of these concepts, consider the following examples:
A \emph{verifier} is an analyzer of type~$P \times \Bspec \to R \times \Omega$, 
which takes as input a program~$p$ and a behavior specification~$\bspec$,
and produces as output a result~$r$ and a witness~$\omega$\,%
\footnote{All verifiers that participate in the competition SV-COMP are analyzers of this form.}.
A \emph{conditional verifier} is of type~$P \times \Bspec \times \Psi \to R \times \Omega \times \Psi$,
i.e., a verifier that supports also input and output conditions.
A \emph{validator} is of type~$P \times \Bspec \times \Omega \to R \times \Omega$,
i.e., a verifier that takes as input in addition a witness.
A \emph{test-case generator} is also an analyzer, but of type~$P \times \Tspec \to 2^T$,
which takes as input a program $p$ and a test specification~$\tspec$,
and produces as output a set $ts \in 2^T$ of test cases. 

\iftechreport
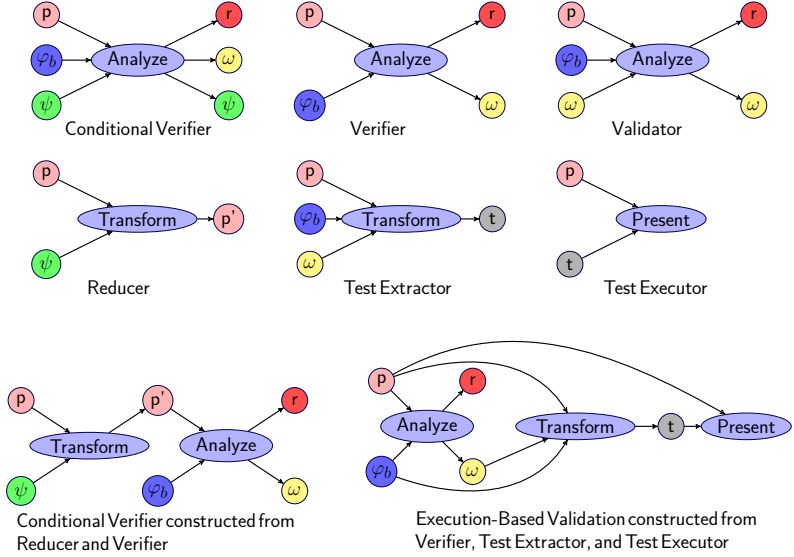
\begin{figure}[t]
\centering
\scalebox{0.3}{
    %Conditional Verifier
    \begin{tikzpicture}[node distance = 10 mm and 10 mm,> = stealth']
    \node (a) [cloud] at (-10,0) {\Huge Analyze};
    \node (l) [text width=7cm] at (-9.7,-3) {\Huge Conditional Verifier};
    \node (p) [circle-P] at (-14,2) {\Huge p};
    \node (phi) [circle-phi] at (-14,0) {\Huge $\bspec$};
    \node (psi) [circle-psi] at (-14,-2) {\Huge $\psi$};
    \node (r) [circle-r] at (-6,2) {\Huge r};
    \node (w) [circle-w] at (-6,0) {\Huge $\omega$};
    \node (psi2) [circle-psi] at (-6,-2) {\Huge $\psi$};
    \draw[->, line width=.5mm] (a) -- (r);
    \draw[->, line width=.5mm] (a) -- (w);
    \draw[->, line width=.5mm] (a) -- (psi2);
    \draw[->, line width=.5mm] (p) -- (a);
    \draw[->, line width=.5mm] (phi) -- (a);
    \draw[->, line width=.5mm] (psi) -- (a);
    \end{tikzpicture}
    \hspace{20mm}
    %Verifier
    \begin{tikzpicture}[node distance = 10 mm and 10 mm,> = stealth']
    \node (a) [cloud] at (-10,0) {\Huge Analyze};
    \node (l) [text width=5cm] at (-9.7,-3) {\Huge Verifier};
    \node (p) [circle-P] at (-14,2) {\Huge p};
    \node (phi) [circle-phi] at (-14,-2) {\Huge $\bspec$};
    \node (r) [circle-r] at (-6,2) {\Huge r};
    \node (w) [circle-w] at (-6,-2) {\Huge $\omega$};
    \draw[->, line width=.5mm] (a) -- (r);
    \draw[->, line width=.5mm] (a) -- (w);
    \draw[->, line width=.5mm] (p) -- (a);
    \draw[->, line width=.5mm] (phi) -- (a);
    \end{tikzpicture}
    \hspace{20mm}
    %Validator
    \begin{tikzpicture}[node distance = 10 mm and 10 mm,> = stealth']
    \node (a) [cloud] at (-10,0) {\Huge Analyze};
    \node (l) [text width=5cm] at (-9.7,-3) {\Huge Validator};
    \node (p) [circle-P] at (-14,2) {\Huge p};
    \node (phi) [circle-phi] at (-14,0) {\Huge $\bspec$};
    \node (w) [circle-w] at (-14,-2) {\Huge $\omega$};
    \node (r) [circle-r] at (-6,2) {\Huge r};
    \node (w2) [circle-w] at (-6,-2) {\Huge $\omega$};
    \draw[->, line width=.5mm] (a) -- (r);
    \draw[->, line width=.5mm] (a) -- (w2);
    \draw[->, line width=.5mm] (p) -- (a);
    \draw[->, line width=.5mm] (phi) -- (a);
    \draw[->, line width=.5mm] (w) -- (a);
    \end{tikzpicture}
}\\[2ex]
\scalebox{0.3}{
    %Reducer
    \begin{tikzpicture}[node distance = 10 mm and 10 mm,> = stealth']
    \node (a) [cloud] at (-10,0) {\Huge Transform};
    \node (l) [text width=5cm] at (-9.7,-3) {\Huge Reducer};
    \node (p) [circle-P] at (-14,2) {\Huge p};
    \node (psi) [circle-psi] at (-14,-2) {\Huge $\psi$};
    \node (p2) [circle-P] at (-6,0) {\Huge p'};
    \draw[->, line width=.5mm] (a) -- (p2);
    \draw[->, line width=.5mm] (p) -- (a);
    \draw[->, line width=.5mm] (psi) -- (a);
    \end{tikzpicture}
    \hspace{20mm}
    %Test-Extractor
    \begin{tikzpicture}[node distance = 10 mm and 10 mm,> = stealth']
    \node (a) [cloud] at (-10,0) {\Huge Transform};
    \node (l) [text width=7cm] at (-9.0,-3) {\Huge Test Extractor};
    \node (p) [circle-P] at (-14,2) {\Huge p};
    \node (phi) [circle-phi] at (-14,0) {\Huge $\bspec$};
    \node (w) [circle-w] at (-14,-2) {\Huge $\omega$};
    \node (t) [circle-t] at (-6,0) {\Huge t};
    \draw[->, line width=.5mm] (a) -- (t);
    \draw[->, line width=.5mm] (p) -- (a);
    \draw[->, line width=.5mm] (phi) -- (a);
    \draw[->, line width=.5mm] (w) -- (a);
    \end{tikzpicture}
    \hspace{20mm}
    %Test-Executor
    \begin{tikzpicture}[node distance = 10 mm and 10 mm,> = stealth']
    \node (a) [cloud] at (-10,0) {\Huge Present};
    \node (l) [text width=7cm] at (-9.0,-3) {\Huge Test Executor};
    \node (p) [circle-P] at (-14,2) {\Huge p};
    \node (t) [circle-t] at (-14,-2) {\Huge t};
    \draw[->, line width=.5mm] (p) -- (a);
    \draw[->, line width=.5mm] (t) -- (a);
    \end{tikzpicture}
}\\[1ex]
\scalebox{0.3}{
    %Construct Conditional Verifier from Verifier
    \begin{tikzpicture}[node distance = 10 mm and 10 mm,> = stealth']
    \node (a) [cloud] at (-21,0) {\Huge Transform};
    \node (b) [cloud] at (-15,0) {\Huge Analyze};
    \node (l) [text width=14cm] at (-17.2,-3.8) {\Huge Conditional Verifier constructed from \\[1ex]
                                                       Reducer and Verifier};
    \node (p) [circle-P] at (-24,2) {\Huge p};
    \node (psi) [circle-psi] at (-24,-2) {\Huge $\psi$};
    \node (p2) [circle-P] at (-18,2) {\Huge p'};
    \node (phi) [circle-phi] at (-18,-2) {\Huge $\bspec$};
    \node (r) [circle-r] at (-12,2) {\Huge r};
    \node (w) [circle-w] at (-12,-2) {\Huge $\omega$};
    \draw[->, line width=.5mm] (a) -- (p2);
    \draw[->, line width=.5mm] (p) -- (a);
    \draw[->, line width=.5mm] (psi) -- (a);
    \draw[->, line width=.5mm] (p2) -- (b);
    \draw[->, line width=.5mm] (phi) -- (b);
    \draw[->, line width=.5mm] (b) -- (r);
    \draw[->, line width=.5mm] (b) -- (w);
    \end{tikzpicture}
    \hspace{10mm}
    %Executing-Based Validation after Verification
    \begin{tikzpicture}[node distance = 10 mm and 10 mm,> = stealth']
    \node (a) [cloud] at (-22,0) {\Huge Analyze};
    \node (b) [cloud] at (-15.5,0) {\Huge Transform};
    \node (c) [cloud] at (-8,0) {\Huge Present};
    \node (l) [text width=16cm] at (-14.5,-4.6) {\Huge Execution-Based Validation constructed from \\[1ex]
                                                       Verifier, Test Extractor, and Test Executor};
    \node (p) [circle-P] at (-24,2) {\Huge p};
    \node (phi) [circle-phi] at (-24,-2) {\Huge $\bspec$};
    \node (r) [circle-r] at (-20,2) {\Huge r};
    \node (w) [circle-w] at (-20,-2) {\Huge $\omega$};
    \node (t) [circle-t] at (-11.3,0) {\Huge t};
    \draw[->, line width=.5mm] (p) -- (a);
    \draw[->, line width=.5mm] (phi) -- (a);
    \draw[->, line width=.5mm] (p) to [out=30,in=150] (c);
    \draw[->, line width=.5mm] (phi) to [out=-20,in=-120]  (b);
    \draw[->, line width=.5mm] (p) to [out=20,in=120]  (b);
    \draw[->, line width=.5mm] (a) -- (r);
    \draw[->, line width=.5mm] (a) -- (w);
    \draw[->, line width=.5mm] (w) -- (b);
    \draw[->, line width=.5mm] (b) -- (t);
    \draw[->, line width=.5mm] (t) -- (c);
    \end{tikzpicture}
}
  \caption{Graphical visualization of the component framework}
  \label{fig:component-framework}
\end{figure}
\fi

Transformers are largely lacking today, only a few exist already~\cite{Reducer,ExecutionBasedValidation}.
Transformers are, however, key to cooperation: 
only if a transformer can bring the artifact into a form 
understandable by the next tool without implementing an extension of this tool, 
cooperation can be put into practice. 
A \emph{test-case extractor} is a transformer of type~$P \times \Bspec \times \Omega \to T$,  
which translates a program, specification, and violation witness to a test case.
The identity function is also a transformer (for any given type).
A \emph{reducer} is a transformer of type~$P \times \Psi \to P$,
which takes a program and a condition as input, and transforms it to a residual program.
% More:
%\testtowitness is a transformer~$T \to W$,
%which takes as input a test case from~$T$
%and produces as output a witness from~$W$.
%\testselector is a transformer~$(\testset, S) \to T$,
%which takes as input a test set from~$\testset$ and a specification from~$S$,
%and produces as output a test case~$T$ that violates the specification.

Presenters form the interface to the user.
A \emph{test-case executor} is a presenter of type~$P \times T \to \{\}$,
which takes a program~$p$ and a test case $T$ as input,
and shows a particular program execution to the software engineer. 
% Alternative definition:
%\testexecutor is a transformer~$(T, P) \to (R, D)$,
%which takes as input a test case from~$T$ and a program from~$P$,
%and produces as output a result from~$R$ and a stack trace from~$D$.

\newcommand{\redu}{\mathit{red}}
\newcommand{\veri}{\mathit{ver}}
\newcommand{\wtot}{\mathit{wit2test}}
\newcommand{\exe}{\mathit{exec}}
Now we can construct, for example, a conditional verifier from a reducer~$\redu$ and an off-the-shelf verifier~$\veri$ by composition.
For inputs $p$ and $\bspec$, the expression $\veri(\redu(p, \bspec), \bspec)$ runs the construction.
For a verification with an execution-based result validation based on a given verifier~$\veri$, test extractor~$\wtot$, and test executor~$\exe$,
we can write $\exe(p, \wtot(p, \bspec, \veri(p, \bspec).\omega))$.
\iftechreport
  \Cref{fig:component-framework} 
  shows a graphical visualization of the individual components
  and the two mentioned constructions.
\else
  The extended version of this article~\cite{CooperativeVerification-TR} shows a graphical visualization
  of the components and the two constructions.
\fi

With our construction framework, it is possible to identify the gaps of meaningful transformers,
and propose solutions to close these gaps,  
as far as needed for cooperation.

%\emph{Combinations.}
%The systematic classification of tools (plus the additional implementation of new transformers) 
%opens up the way to new combinations of tools. 
%Consider for instance the use case that a developer without knowledge of verification technology
%wants to use a model-checking tool to find a bug regarding specification~$S$ in program~$P$.
%The developer can use off-the-shelf components
%$\textsf{v} \in \verifier$, $\textsf{w} \in \witnesstotest$, and $\textsf{t} \in \testexecutor$,
%and assemble $\textsf{v} \circ \textsf{w} \circ \textsf{t}$, which is an analyzer mapping $(P, S, P)$ to $(R, R, D)$. \hw{nur ein P?}
%By running the resulting transformer on the program and the specification,
%the developer obtains a stack trace~$D$ that can be inspected with a debugger of the developer's choice.
%
%
%
%\begin{figure}[t]
%  Drawings of some configurations
%  \caption{Usecases in the developer's work flow}
%\end{figure}

%\input{cooperation-design}

\subsection{Semantics} \label{sec:semantics}
We now develop the theoretical foundations of artifacts and actors. 
Artifacts describe some information about a program (or a program itself), and 
for sound cooperation we need to define the \emph{semantics} of artifacts. 
For instance, a violation witness of a program describes a path of the program on 
which a specific specification is violated, a condition  
describes a set of paths of a program which have (or have not been) inspected by an analyzer. 
When employing cooperation as a means for sharing the work load of validation, 
the cooperating tools need to agree on the meaning of the exchanged artifacts. 
Without this, cooperation might easily get unsound, e.g., returning a result \true for 
a program and specification although the combined usage of tools has failed to inspect 
the whole state space of the program.  
By defining the semantics of artifacts, we also implicitly define the desired semantics of 
the various actors operating on artifacts.

% TODO\\
% - add citations
\nocite{COVERITEST,CompactProofWitnesses,CzechConditionalModelChecking,JakobsCertificateReductionPartitioning,ChristakisMW16} 

\begin{figure}[t] 
\myvspace{-3mm}
\centering 
\begin{subfigure}[b]{.3\textwidth}
\begin{verbatim}
  0:  int x = input(); 
  1:  int a = 0;
  2:  int b = 0;
  3:  while (a < x) {
  4:       a++;
  5:       b++;   // later elided  
       }
  6:  


\end{verbatim} 
\caption{Example program \prog}
\end{subfigure} \hfill
\begin{subfigure}[b]{.5\textwidth}
\centering
\begin{tikzpicture}[node distance = .55cm, initial text={},auto,shorten >=1pt,scale=0.8, every node/.style={scale=0.8}]
\node[ca] (i) {$0$};
\node[ca]  (0) [right=of i,xshift=2cm] {$1$};
\node[ca] (1) [below=of 0] {$2$};
\node[ca] (2) [below=of 1] {$3$};
\node[ca] (3) [below left=of 2,xshift=-.3cm] {$4$};
\node[ca] (4) [below=of 3] {5}; 
\node[ca] (6) [below right=of 2, xshift=.3cm] {$6$};
\path[-stealth]  (i) edge node [yshift=2mm]{\textcolor{blue}{(0,int x = input(), 1)}} (0);
\path[-stealth]  (0) edge node {\textcolor{blue}{(1,int a=0, 2)}} (1);
\path[-stealth]  (1) edge node {\textcolor{blue}{(2, int b=0, 3)\ }} (2);
\path[-stealth]  (2) edge node [left,yshift=2mm] {\textcolor{blue}{\ (3, a<x, 4)}} (3);
\path[-stealth]  (2) edge node [right,yshift=2mm] {\textcolor{blue}{\ (3, !(a<x), 6)}} (6);
\path[-stealth]  (3) edge node [left] {\textcolor{blue}{(4, a++, 5)\ }} (4);
\path[-stealth]  (4) edge node [right,xshift=-3mm,yshift=-6mm] {\textcolor{blue}{(5, b++, 3)\ }} (2);
\end{tikzpicture}
\caption{Control-flow automaton~$A_p$}
\end{subfigure}
\myvspace{-3mm}
\caption{Example program and its control-flow automaton}
\label{fig:exprog}
\myvspace{-8mm}
\end{figure}
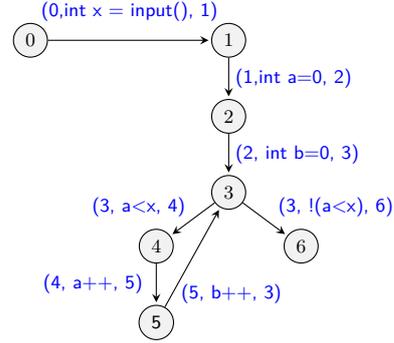 

We start the formalization of artifacts with the definition of programs, our prime artifact.  
We denote the set of all program locations by $Loc$.  Formally, 
a program~\prog is described by a {\em control-flow automaton} (CFA) $A_p=(L,\ell_0,G)$ 
that consists of a set of locations $L \subseteq Loc$, 
an initial location $\ell_0 \in L$, and a set of control-flow edges $G \subseteq L \times Ops \times L$,
where $Ops$ is the set of operations. Operations can be (a) assignments,   
(b) assume statements 
(arising out of branches), and (c) calls to functions retrieving inputs. 
Here, we assume to have a single such function, called \texttt{input}.  
We let  $\edges = L \times Ops \times L$ 
be the set of all control-flow edges. 
  
We let $X$ be the set of variables occurring in the operations $Ops$.  
For simplicity, we restrict the type of variables to integers. 
A \emph{concrete data state}~$c: X \pto \mathbb{Z}$ is thus a partial mapping from $X$ to $\mathbb{Z}$.
% An input to $P$ (or test case of $P$) is a mapping $tc: \Inp \rightarrow \mathbb{Z}$. 
In the left of \cref{fig:exprog} we see our running example of the simple program \prog
and its control-flow automaton on the right. The program starts by retrieving an input 
for variable $x$, sets variables $a$ and $b$ to 0, and then increments both while the value of $a$ is less
than that of $x$. 

A {\em concrete program path} of a program $A_p=(L,\ell_0,G)$  
is a sequence 
$(c_0,\ell_0) \trans {g_1} \ldots \trans {g_n}(c_n,\ell_n)$, where the initial
concrete data state $c_0 = \emptyset$ assigns no value,
$g_i = (\ell_{i-1},op_i,\ell_i) \in G$, and $c_{i-1} \trans{op_i} c_i$, i.e.,
(a) in case of assume operations, $c_{i-1} \models op_i$ ($op_i$~is a boolean condition) 
and $c_{i-1} = c_i$,
(b) in case of assignments, ${c_i = \mathsf{SP}_{op_i}(c_{i-1})}$, where $\mathsf{SP}$ 
is the strongest-post operator of the operational semantics,
and (c) in case of inputs of the form $x=$\texttt{ input()}, $c_i(x) \in \mathbb{Z}$ 
(nondeterministic choice of input) and $c_i(y) = c_{i-1}(y)$ for all $y \neq x$. 
%and (c) in case of method calls, $c_{i-1} = c_i$. 
An edge $g$ is {\em contained} in a concrete program path 
$\pi = (c_0,\ell_0) \trans {g_1} \ldots \trans {g_n}(c_n,\ell_n)$ if $g = g_i$ for some $i$, $1 \leq i \leq n$.  
%a set of edges $E \subseteq \edges$ is contained in a set of concrete paths $\Pi$ if there exists 
%a $\pi \in \Pi$ for every $g \in E$ such that $g$ is contained in $\pi$. 
%A concrete program path $\pi=(c_0,\ell_0) \trans {g_1} \ldots \trans {g_n}(c_n,\ell_n)$ implies  
%an {\em execution} $\exec(\pi) = c_0 c_1 \ldots c_n$.
We let $\pth(A_p)$ be the set of all concrete program paths.
% and $\exec(P)$ be the set of executions of a CFA~$C$.

We allow artifacts to state {\em assumptions} on program variables. 
These are given as state conditions (from a set $\Pred$ of predicates over a certain theory).
We write $c \models \pred$ to say that a concrete state~$c$ satisfies a 
state condition $\pred \in \Pred$.
%
%\medskip
%\noindent Artifacts 
%\begin{enumerate}
%  \item programs,
%  \item specifications,
%  \item test inputs,
%  \item (error or correctness) witnesses, and 
%  \item conditions 
%\end{enumerate} 
%
%Furthermore: results. 
%
Artifacts on a program $\prog$ are represented by artifact automata: 

\begin{definition} \label{def:aut}
    An \textbf{artifact automaton}  $A = (Q,\Sigma,\delta,q_0,\Inv,F)$ for a program CFA~$A_p=(L,\ell_0,G)$ consists of  
     \begin{itemize}
         \item a finite set $Q$ of {\em states} and an initial state $q_0 \in Q$,
         \item an {\em alphabet} $\Sigma \subseteq 2^G \times \Pred$,
         \item a {\em transition relation} $\delta \subseteq Q \times \Sigma \times Q$, 
         \item an {\em invariant mapping} $\Inv: Q \rightarrow \Pred$, and 
         \item a set $F \subseteq Q$ of {\em final states}.  
     \end{itemize} 
\end{definition}

We write $q \trans {(D,\pred)} q'$ for $(q,(D,\pred),q') \in \delta$. 
In figures, we often elide invariants as well as the assumptions on edges when they are $\mathit{true}$.
We furthermore elide the set notation when the element of $2^G$ is a singleton. 

Artifact automata describe paths of a program. Depending on the sort of artifact automaton, 
these could for instance be paths allowed or disallowed 
by a specification, or paths already checked by a verifier. 
A path of the program can be {\em accepted} (if the automaton 
reaches a final state) or {\em covered} by the automaton.

\begin{definition} \label{def:cover}
  An artifact automaton $A = (Q,\Sigma,\delta,q_0,\Inv,F)$ \textbf{matches} a path 
      $\pi = (c_0,\ell_0) \trans {g_1} (c_1,\ell_1) \trans {g_2} \ldots \trans{g_n} (c_n,\ell_n)$ 
      if there is a run $\rho = q_0 \trans {(G_1,\pred_1)} q_1 \trans {(G_2,\pred_2)} 
      \ldots \trans {(G_k,\pred_k)} q_k, 0 \leq k \leq n$, in $A$, s.t.\
   \begin{enumerate}
       \item $\forall i, 1 \leq i \leq k: g_i \in G_i$, 
       \item $\forall i, 0 \leq i \leq k: c_i \models \Inv(q_i)$ and
       \item $\forall i, 1 \leq i \leq k: c_i \models \pred_i$.
   \end{enumerate}
   The artifact automaton $A$ \textbf{accepts} the path~$\pi$ if $A$ matches $\pi$ and $q_k \in F$, and
   \linebreak
   $A$ \textbf{covers} $\pi$ if $A$ matches $\pi$ and $k = n$.
\end{definition}

We let $L(A)$ be the set of paths accepted by the automaton $A$ (its {\em language}) 
and $\pth(A)$ be the set of paths covered by $A$. 
As we will see below, some artifact automata might have an empty set of final states 
and just des\-cribe a set of paths that they cover. 
 
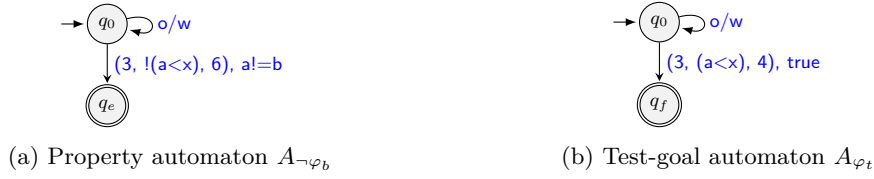
\begin{figure}[t] 
\centering 
\begin{subfigure}[b]{.4\textwidth}
\centering
 \begin{tikzpicture}[node distance = .55cm, initial text={},auto,shorten >=1pt,scale=0.8, every node/.style={scale=0.8}]
\node[ca,initial]  (0) {$q_0$};
\node[caf] (1) [below=of 0] {$q_e$};
\path[-stealth]  (0) edge [loop right] node {\textcolor{blue}{o/w}} (1);
\path[-stealth]  (0) edge node {\textcolor{blue}{(3, !(a<x), 6), a!=b\ }} (1);
\end{tikzpicture}
\caption{Property automaton~$A_{\lnot\bspec}$}
\label{fig:prop}
\end{subfigure} \hfill
\begin{subfigure}[b]{.4\textwidth}
\centering
\begin{tikzpicture}[node distance = .55cm, initial text={},auto,shorten >=1pt,scale=0.8, every node/.style={scale=0.8}]
\node[ca,initial]  (0) {$q_0$};
\node[caf] (1) [below=of 0] {$q_f$};
\path[-stealth]  (0) edge [loop right] node {\textcolor{blue}{o/w}} (1);
%\path[-stealth]  (0) edge node [left] {\textcolor{blue}{(3, a<x, 4), true}} (1);
\path[-stealth]  (0) edge node [right] {\textcolor{blue}{(3, (a<x), 4), true \ }} (1);
\end{tikzpicture}
\caption{Test-goal automaton~$A_{\tspec}$}
\label{fig:tga}
\end{subfigure}
\myvspace{-3mm}
\caption{Automata for a property and a test-goal specification}
\label{fig:spec}
\myvspace{-6mm}
\end{figure}

\medskip
\inlineheadingbf{Artifact Automata as Representation of Artifacts}
We consider different specializations of artifact automata and
use the notation~$A_s$ to denote the automaton that
represents the syntactical object~$s$.
%The first two are employed for specification purposes. 

\medskip
\noindent
\textbf(1)
A \textbf{property automaton} (or, observer automaton) $A_{\lnot\bspec} = (Q,\Sigma,\delta,q_0,\Inv,F)$ is an artifact automaton
         that satisfies the following conditions:  
         \begin{enumerate}
             \item \fbox{$\forall q\in Q: \Inv(q) = true$}, 
	       \item \fbox{$\forall q \in Q\setminus F, \forall g \in G:
                    \bigvee \left\{ \pred~\mid~\exists q' , \exists D, g \in D: q \trans {(D,\pred)} q' \right\} = true$}\\   
             (assuming $\bigvee \emptyset = \mathit{false}$). 
        \end{enumerate} 
         Condition 2 ensures that property automata only observe the state of the program 
         (when running in parallel with the program). They do not block, 
         except for the case when the final state is reached where blocking is allowed. 
         Final states denote the reaching of property violations (or, targets). 

\medskip
\noindent
\textbf(2)
A \textbf{test-goal automaton} $A_{\tspec} = (Q,\Sigma,\delta,q_0,\Inv,F)$ is an artifact automaton
         that has only trivial state invariants, i.e., \fbox{$\forall q \in Q: \Inv(q) = true.$} 
        %\hw{auch die bedingung 2 von oben mit keiner Restriktion???}\db{Nein, kann ruhig engeschraenkt werden.}
%        A test goal automaton for goals $\goals \subseteq G$ is 
%        an artifact automaton with $Q=\{q_0,q_f\}$, $F=\{q_f\}$, $\Inv(q) = true$ for all 
%        $q\in Q$, $q_0 \trans {(g,true)} q_0$ for all $g \in G \setminus \goals$ and 
%        $q_0 \trans {(g,true)} q_e$ for $g \in \goals$. \\ 
        If a final state is reached, the test goal is fulfilled. 

\Cref{fig:spec} shows two specification automata: 
In \cref{fig:prop} we see a property automaton specifying that variables $a$ and $b$ 
have to be equal when the loop terminates, i.e., the error state is reached if 
there is a transition from location 3 to 6 at which $a \neq b$. 
The label \texttt{o/w} (otherwise) denotes all transitions other than the ones 
explicitly depicted.
\Cref{fig:tga} depicts a test-goal automaton for the branch condition 
entering the loop.  

%\smallskip
%\noindent The second sort of artifact automata are {\em witnesses} which witness some facts computed about 
%a program. 

\medskip
\noindent
\textbf(3)
A \textbf{violation-witness automaton} $A_\omega = (Q,\Sigma,\delta,q_0,\Inv,F)$ is again an artifact automaton
            with trivial state invariants only, i.e., \fbox{$\forall q \in Q: \Inv(q) = true.$} 

        Violation witnesses are used to describe the part of a program's state space which contains the error. 
        The final state is reached if an error is detected. 
        Counter examples are a specific form of violation witnesses which describe a single path. 

\medskip
\noindent
\textbf(4)
A \textbf{correctness-witness automaton} $A_\omega = (Q,\Sigma,\delta,q_0,\Inv,F)$ is an artifact automaton 
        that has only trivial transition assumptions, i.e., 
           \fbox{$\forall (q,(D,\pred),q') \in \delta: \pred=true$}, and there are no final states (\fbox{$F=\emptyset$}). 
         A correctness witness typically gives information about the state space of the program  (like a loop invariant) in order to 
         facilitate its verification.      

\begin{figure}[t] 
\centering 
\begin{subfigure}[b]{.45\textwidth}
\centering
 \begin{tikzpicture}[node distance = .55cm, initial text={},auto,shorten >=1pt,scale=0.8, every node/.style={scale=0.8}]
\node[ca,label=right:true,initial] (i) {$q_0$}; 
\node[ca,label=right:true]  (0) [below=of i] {$q_1$};
\node[ca,label=right:{$a=0$}] (1) [below=of 0] {$q_2$};
\node[ca,label=right:{$a=b$}] (2) [below=of 1] {$q_3$};
\node[ca,label=left:{$a=b$}] (3) [below left=of 2,xshift=-.3cm] {$q_4$};
\node[ca,label=left:{$a-1=b$}] (4) [below=of 3] {$q_5$}; 
\node[ca,label=right:{$a=b$}] (6) [below right=of 2, xshift=.3cm] {$q_6$};
\path[-stealth]  (i) edge node {\textcolor{blue}{(0,int x=input(), 1)}} (0);
\path[-stealth]  (0) edge node {\textcolor{blue}{(1,int a=0, 2)}} (1);
\path[-stealth]  (1) edge node {\textcolor{blue}{(2, int b=0, 3)\ }} (2);
\path[-stealth]  (2) edge node [left,yshift=1mm] {\textcolor{blue}{\ (3, a<x, 4)}} (3);
\path[-stealth]  (2) edge node [right,yshift=1mm] {\textcolor{blue}{\ (3, !(a<x), 6)}} (6);
\path[-stealth]  (3) edge node [left] {\textcolor{blue}{(4, a++, 5)\ }} (4);
\path[-stealth]  (4) edge node [right,xshift=-3mm,yshift=-6mm] {\textcolor{blue}{(5, b++, 3)\ }} (2);
\end{tikzpicture}
\caption{Correctness witness automaton $A_\omega$}
\label{fig:corr-witness}
\end{subfigure} \hfill
\begin{subfigure}[b]{.45\textwidth}
\centering
\begin{tikzpicture}[node distance = .5cm, initial text={},auto,shorten >=1pt,scale=0.8, every node/.style={scale=0.8}]
\node[ca,initial] (i) {$q_0$}; 
\node[ca]  (0) [below=of i] {$q_0$};
\node[ca] (1) [below=of 0] {$q_1$};
\node[ca] (2) [below=of 1] {$q_2$};
\node[ca] (3) [below=of 2] {$q_3$};
\node[ca] (4) [below=of 3] {$q_4$}; 
\node[caf] (5) [below=of 4] {$q_e$};
\path[-stealth]  (i) edge node {\textcolor{blue}{(0,int x=input(), 1), $x\geq 1$}} (0);
\path[-stealth]  (0) edge node {\textcolor{blue}{(1,int a=0, 2)}} (1);
\path[-stealth]  (1) edge node {\textcolor{blue}{(2, int b=0, 3)\ }} (2);
\path[-stealth]  (2) edge node {\textcolor{blue}{\ (3, a<x, 4)}} (3);
\path[-stealth]  (3) edge node {\textcolor{blue}{\ (4, a++, 3)}} (4);
\path[-stealth]  (4) edge node {\textcolor{blue}{(3, !(a<x), 6)\ }} (5);
\end{tikzpicture}
\caption{Violation witness automaton $A_\omega$}
\label{fig:error-witness}
\end{subfigure}
\myvspace{-3mm}
\caption{Automata for a correctness witness for program \prog and a violation witness for \prog without line 5, 
   both wrt.~behavior specification $\bspec$ of \cref{fig:prop}}
\label{fig:witnesses}
\vspace{-4mm}
\end{figure}
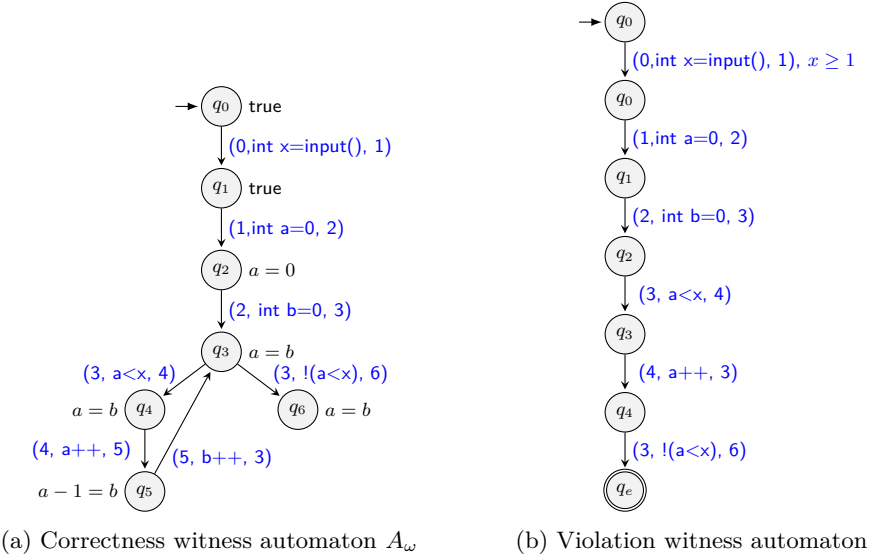

In \cref{fig:witnesses} we see both a correctness and a violation witness. 
The correctness witness belongs to program \prog and, e.g.,~certifies that at location 3 variables $a$ and $b$ are equal 
(via the invariant for $q_3$). 
The violation witness on the right belongs to program \prog with line 5 removed, i.e., a program which does not satisfy the 
property stated in \cref{fig:prop}. 
The violation witness states that an input value of $x$ being greater or equal to  1 is needed for directing the 
verifier towards the error. 

%\smallskip
%\noindent 
%The next two automata are neither witnesses nor specification automata. 

\medskip
\noindent
\textbf(5)
A \textbf{condition automaton} $A_\psi = (Q,\Sigma,\delta,q_0,\Inv,F)$ is an artifact automaton 
       that satisfies 
       \begin{enumerate} 
           \item \fbox{$\forall q \in Q: \Inv(q) = true$} (no invariants on states) and 
           \item \fbox{$\neg \exists (q_f,*,q) \in \delta$ with $q_f \in F$}  
              (no transitions leaving final states). %\hw{soll diese Bedingung auch irgendwo an die anderen Automaten?}
      \end{enumerate} 
      A condition is typically used to describe parts of the state space of a program, e.g.,~the part 
      already explored during verification. 
     Final states are thus used to fix which paths have already been explored. 

A test case is a sequence of input values 
       consecutively supplied to the calls of function \texttt{input}. 
       Such a test case is encoded as 
       artifact automaton using a special template variable~$\x$ that can be instantiated with 
       every program variable.  

\medskip
\noindent
\textbf(6)
A \textbf{test-case automaton} $A_t = (Q,\Sigma,\delta,q_0,\Inv,F)$ for a test case $\langle z_1, \ldots, z_n \rangle$  
       is an artifact automaton with the following components:
       \fbox{$Q = \{q_0, \ldots, q_n\}$, $q_i \trans {((*,\x=\texttt{input()},*),\x=z_{i+1})} q_{i+1}$, 
       $q_i \trans {\texttt{o/w}} q_{i}$} ($0 \leq i < n$) 
       and \fbox{$F=\emptyset$}.  
       %For these special labels on the transitions we need to extend our coverage definition. 
       For matching these special transitions $(G_i,\pred_i) = ((*,\x=\texttt{input()},*),\x=z)$ with program paths, 
       the program transitions $g_i$ have to be of the form $(\ell,x=\texttt{input()},\ell')$ and the next state needs to satisfy $c_i(x)=z$, 
      $c_i(y) = c_{i-1}(y)$ for $y\neq x$.  

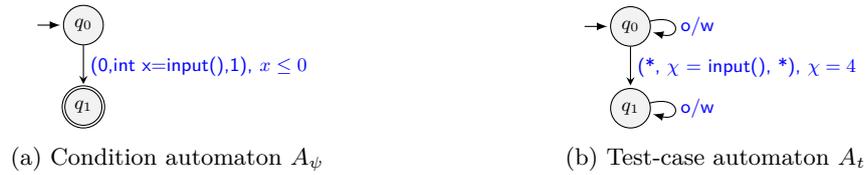
\begin{figure}[t] 
\centering 
\begin{subfigure}[b]{.4\textwidth}
\centering
\begin{tikzpicture}[node distance = .55cm, initial text={}, auto,shorten >=1pt,scale=0.8, every node/.style={scale=0.8}]
 \node[ca,initial]  (0) {$q_0$};
\node[caf] (1) [below=of 0] {$q_1$};
%\node[caf] (2) [below=of 1] {$q_f$};
\path[-stealth]  (0) edge node {\textcolor{blue}{(0,int x=input(),1), $x\leq 0$}} (1);
%\path[-stealth]  (1) edge node {\textcolor{blue}{(2, int b=0, 3), $x\leq 0$}} (2);
\end{tikzpicture}
\caption{Condition automaton $A_\psi$}
\label{fig:condition}
\end{subfigure} \hfill
\begin{subfigure}[b]{.4\textwidth}
\centering
\begin{tikzpicture}[node distance = .55cm, initial text={},auto,shorten >=1pt,scale=0.8, every node/.style={scale=0.8}]
 \node[ca,initial]  (0) {$q_0$};
\node[ca] (1) [below=of 0] {$q_1$};
%\path[-stealth] (0) edge [loop right] node {\textcolor{blue}{o/w}} (0);
\path[-stealth]  (0) edge [loop right] node {\textcolor{blue}{o/w}} (0);
\path[-stealth]  (0) edge node {\textcolor{blue}{(*, $\x$ = input(), *), $\x = 4$}} (1);
\path[-stealth]  (1) edge [loop right] node {\textcolor{blue}{o/w}} (1);
\end{tikzpicture}
\caption{Test-case automaton $A_t$}
\label{fig:test-vector}
\end{subfigure}
\myvspace{-2mm}
\caption{Automata for a condition and a test case} 
\label{fig:ctv}
\myvspace{-6mm}
\end{figure}

 \Cref{fig:condition} gives a condition stating the exploration of the state space for inputs less or equal to 0. 
This could for instance be the output of a verifier having checked that the property holds for inputs $x \leq 0$. 
\Cref{fig:test-vector} is the test-case automaton for the test case~$\langle 4 \rangle$. 

\medskip
\inlineheadingbf{Semantics of Artifact Automata}
The above definitions fix the syntactical structure of artifact automata. 
In addition, we need to state their semantics, i.e., the meaning of particular artifacts for a given program.
In the following, we let $A_p=(L,\ell_0,G)$ be the CFA for a program~$\prog$
and $A_{\lnot\bspec}$, $A_\omega$, and $A_{\tspec}$ be artifact automata. 

\medskip
\noindent
\textbf{(i)} The program $\prog$ {\em fulfills} a property specification~$\bspec$
         if 
         \fbox{$\pth(A_p) \cap L(A_{\lnot\bspec}) = \emptyset$}.  
     Our running example \prog fulfills the property of \cref{fig:prop}.   
  
\medskip
\noindent
\textbf{(ii)} A correctness witness $\omega$ is {\em valid} for a program $\prog$ and property specification~$\bspec$
        if 
        \fbox{$\pth(A_p) \subseteq \pth(A_\omega) \wedge \pth(A_p) \cap L(A_{\lnot\bspec}) = \emptyset$}.
    We see here that a correctness witnesses can thus be used to facilitate verification: 
    when we run program, property, and correctness witness in parallel in order to check  the emptiness of 
    $\pth(A_p) \cap L(A_{\lnot\bspec})$, the correctness witness helps in proving the program correct. 
    The correctness witness in \cref{fig:corr-witness} is valid for \prog 
    and the property in \cref{fig:prop}. 
 
\medskip
\noindent
\textbf{(iii)}  A violation witness $\omega$ is {\em valid} for a program $\prog$ and a property specification~$\bspec$
        if 
        \fbox{$\pth(A_p) \cap L(A_\omega) \cap L(A_{\lnot\bspec}) \neq \emptyset$}.
      During verification, violation witnesses can thus steer state-space exploration towards the property violation. 
      Looking again at the running example: If we elide the statement in location~5 of our program, the automaton in 
      \cref{fig:error-witness} is a valid violation witness.
      It restricts the state-space exploration to inputs for variable~$x$ 
      which are greater or equal to~1. 

%\noindent Witnesses always refer to particular properties being proven. 
%Conditions are first of all independent of properties. 

\medskip
\noindent
\textbf{(iv)}  A condition $\psi$ is {\em correct} for a program $\prog$ and property~$\bspec$ 
     if 
     \fbox{$\pth(A_p) \cap L(A_\psi) \cap L(A_{\lnot\bspec}) = \emptyset$}.
     All program paths accepted by the condition fulfill the specification given by the property automaton. 
%     A condition describes a set of paths, namely the ones accepted by the condition. 
%     Typically, a condition is used to fix the part of a program's state space which 
%     has already been inspected  by a tool during verification.
     The condition in \cref{fig:condition} describes all paths of the program $\prog$ 
     which initially started with input $x$ less or equal to 0. 
     This condition is correct for $\prog$ and the property automaton in \cref{fig:prop}. 

%\noindent 
Finally, for testing we are interested in test inputs covering certain test goals. 

\medskip
\noindent
\textbf{(v)} A test-case~$t$ for a program $\prog$ {\em covers} the goals of
       a test-goal specification~$\tspec$ 
       if 
       \fbox{$\pth(A_p) \cap \pth(A_t) \cap L(A_{\tspec}) \neq \emptyset$}.
     Basically, we require that the inputs provided by the test case guarantee program 
     execution to reach (at least one) test goal. If there are more than one final state in the 
     test-goal automaton (or the final state can be reached via different paths), the test-goal 
    automaton specifies several test goals. In this case, the test case covers only some of 
    these goals. %\hw{wollen wir die Definition von test goal automaton so einschränken, dass nur ein goal drin ist?} 
        The test-case automaton in \cref{fig:test-vector} for \prog covers the (single) goal of the test-goal automaton 
    in \cref{fig:tga}. 

\section{Conclusion}

Different verification approaches have different strengths,
and the only way to benefit from a variety of approaches
is to combine them.
The two classic approaches of combining approaches either
in white-box manner via a tight conceptual integration or
in black-bock manner via loosely coupled combinations, such as
portfolio or selection,
are both insufficient.

We propose that the right direction to go is the way of cooperation:
a loosely-coupled combination of tools that interact via clear interfaces and exchange formats,
in order to achieve the verification goal together.
To this end, we provide a classification and 
an overview of existing techniques, which we briefly describe,
while giving most importance to cooperative approaches.

\bibliography{dbeyer,sw,references}
\end{document}